\documentclass{article}
\usepackage{graphicx} 
\usepackage{authblk}
\usepackage[utf8]{inputenc}
\usepackage{geometry}
\usepackage{amssymb}
\usepackage{amsmath}
\usepackage{caption}
\usepackage[makeroom]{cancel}

\usepackage{xcolor}

\usepackage{hyperref}

\title{Reducing Calls to the Simulator in Simulation Based Inference (SBI) }
\author[1]{David Refaeli\thanks{Email: davidrefaeli@tauex.tau.ac.il}}
\author[1]{Mira Marcus-Kalish}
\author[1]{David M. Steinberg}
\affil[1]{Department of Statistics and Operations Research, Tel-Aviv University}
\date{April 2025}

\begin{document}

\maketitle

\begin{abstract}
 Simulation-Based Inference (SBI) deals with statistical inference in problems where the data are generated from a system that is described by a complex stochastic simulator. The challenge for inference in these problems is that the likelihood is intractable; SBI proceeds by using the simulator to sample from the likelihood. In many real world applications, simulator calls are expensive, limiting the associated sample size. Our goal in this work is to extend SBI to exploit two proposals for reducing simulator calls: to draw likelihood samples from a Neural Density Estimator (NDE) surrogate rather than from the stochastic simulator; and use of ``Support Points'' rather than simple random sampling to generate evaluation sites. We embed these methods in the Sequential Neural Posterior Estimator (SNPE) algorithm. Across a suite of test cases, we find that the NDE surrogate improves the quality of the inference; support points worked well in some examples, but not in others.
\end{abstract}

\section{Introduction}

Bayesian inference for model parameters $\theta$ is concerned with finding and exploiting a posterior distribution $p(\theta|x)$ after having obtained data $x$. Bayes formula converts the prior distribution $p(\theta)$ and the likelihood $p(x|\theta)$ into the posterior,
$$p(\theta|x)=\frac{p(x|\theta)p(\theta)}{p(x)}.$$
Obtaining this analytical solution is often hard due to the need to calculate $p(x)=\int_\theta p(x|\theta)p(\theta)d\theta$, the normalizing constant that appears in the denominator. A variety of computational methods can be used to overcome the problem: from simple Monte Carlo methods to Markov Chain Monte Carlo (MCMC) and Variational Inference (VI).
\\

 All the above methods require access to the unnormalized posterior, i.e., the numerator in Bayes formula. In many applications, however, the data generating process is a complex stochastic simulator for which one cannot write the likelihood $p(x|\theta)$ analytically. These are the problems that are addressed by Simulation-based Inference (SBI).
\\

It should be noted that stochastic simulators don't always produce intractable likelihoods. If the model equations are simple enough, analytical solutions (i.e., distributions) exist via stochastic calculus (e.g., Itô’s calculus). But in most real world models, the simulators are too complex for analytical representation.
\\

SBI simulates data samples and uses them as a basis to circumvent the intractable likelihood problem. The initial solutions applied an accept/reject sampling scheme to carry out approximate inference - and were appropriately dubbed Approximate Bayesian Computation (ABC) (\cite{rejabc,mcmcabc,smcabc,toni}). A significant improvement to the original solutions was to model the underlying structure between $x$ and $\theta$, which is then used to improve the performance of the methods. This was originally known as ``Regression Adjustment'' \cite{regadj,blum}. This path grew with the use of newer and more powerful algorithms utilizing artificial neural network-based density estimators \cite{snpe-a,snpe-b,snpe-c,snle-a,snle-b,snre,durkan}. Although these methods can be more powerful than traditional ones, they are less robust to model mis-specification, and one must also weigh in the overhead required to train the neural network.
\\

In many applications, calls to the simulator are expensive. It is thus beneficial to reduce the number of calls needed for inference. In this work we suggest two ways of doing so: using a Neural Density Estimator (NDE) surrogate, and replacing simple random sampling with a sample of ``Support Points'', chosen to better represent the distribution: \\
\begin{itemize}
    \item \textbf{NDE surrogate} - The idea is to model the mapping between $\theta$ and $x$ ($x=f(\theta)$). In deterministic simulators, $f$ is a function that outputs a single value (which can be multi-dimensional), so the mapping might be modeled by a function approximator (neural net, Gaussian process regression, random forest, etc.). For stochastic simulators the mapping results in a 
    conditional distribution $f=p(x|\theta)$. As such we suggest the use of NDE as a fast surrogate (via sampling from the trained NDE) which can replace some of the simulator calls in the Sequential Neural Posterior Estimator (SNPE) algorithm. \\
    \item \textbf{Support Points} - All of the SBI algorithms involve sampling $\theta$ from either a prior or a proposal distribution, and then feeding these points to the simulator. Naïve sampling can be wasteful. We assess, instead, the use of ``Support Points'' \cite{sp}, which lead to a sample that is more ``representative'' of this distribution. \\
\end{itemize} 

\textbf{In section 2} we give an overview of related work in the fields of SBI and surrogates. \textbf{In section 3} we give a detailed view of the new methods used in this work. \textbf{In section 4} we describe the details of how we tested the new methods. \textbf{In section 5} we report the results, and we conclude \textbf{in section 6} with a discussion.
\\

\section{Related Work}

Simulation-based inference (SBI) is employed  to estimate parameters $\theta$ when we have a generating model with built-in randomness, but are unable to write down a density $p(x|\theta)$ for the data. SBI simulates data from the model to compensate for the lack of a density; we will denote such simulation results by $x_{sim}(\theta)$ and denote the observed data by $x_{obs}$. 
\\

Research on SBI began with Approximate Bayesian Computation (ABC), proposed by Tavar\'{e} et al. \cite{rejabc} and further developed by \cite{fu, weiss, pritchard} to address inference problems in computational biology. The rejection ABC method selects $\theta$ from 
an appropriate proposal distribution, generates $x_{sim}(\theta)$, and accepts $\theta$ as a point sampled from the posterior if $x_{sim}(\theta)$ is ``close enough'' to $x_{obs}$. Related ideas include MCMC ABC \cite{mcmcabc} and Sequential Monte Carlo (SMC) ABC \cite{smcabc,toni}. These algorithms all require some ad-hoc decisions, such as choice of a distance measure and a closeness threshold. There is an inherent trade-off between accuracy and efficiency. Tightening the threshold increases accuracy at the price of rejecting more sampled values.
\\

``Regression Adjustment'' \cite{regadj} improves ABC by exploiting the underlying structure connecting $x$ and $\theta$. Given a sample of $\theta$'s and their corresponding $x_{sim}(\theta)$, fit the regression model $\theta = \alpha + x_{sim}\cdot\beta + \xi$.
Then use the regression to adjust all the $\theta$'s to a value more compatible  with $x_{obs}$, $\theta_i^* = \theta_i - x_{sim,i}\cdot \hat \beta + x_{obs}\cdot\hat\beta$. This improves the accuracy-efficiency trade-off, making it possible to accept a much larger fraction of $x_{sim}$ values with little loss in accuracy. The algorithm generalizes to multivariate $\theta$ by using multivariate regression.
\\

The next logical step, to open the regression modeling to non-linear function approximators (e.g., neural networks) and to heteroscedastic noise, was taken by Blum \& François \cite{blum}. Just like in the linear regression, assuming our model captures the true structure, we can adjust the sampled values of $\theta$ to vectors that are more consistent with the observed data.
Another important conceptual modification offered by Blum \& François is to use the algorithm in an adaptive or sequential manner: use the posterior sample generated by a first round of the algorithm to modify the sampling scheme for $\theta$ in the second round, with increased weight in regions that appear to have higher posterior density.
These improvements allowed Blum \& François to increase the threshold and accept up to 90\% of all sample points on some problems.
\\

Subsequent developments utilized Neural Density Estimators (NDEs), which are neural networks that output actual distributions, to discover the complete structure between $x$ and $\theta$, and to obtain a posterior \cite{snpe-a,snpe-b,snpe-c,snle-a,snle-b,snre}. They also implemented sequential sampling, so that the $\theta$'s chosen ``zoom in'' on $p(\theta|x_{obs})$: each round uses the posterior from the last round as the proposal distribution to sample new $\theta$'s. Depending on whether the NN targets the posterior directly, the likelihood, or the likelihood ratio, these algorithms are accordingly called: Neural Posterior Estimator (NPE), Neural Likelihood Estimator (NLE) or Neural Ratio Estimator (NRE). If they are run sequentially an ``S'' is added to the acronym (SNPE, SNLE, SNRE). The non-sequential algorithms try to discover the ``global'' structure (i.e., they find the posterior $p(\theta|x)$ for any $x$ that can be simulated using parameters from the prior). The sequential algorithms concentrate on the ``local'' structure around the observed data, i.e. on $p(\theta|x_{obs})$. The non-sequential algorithms might work well for problems of small complexity, but the sequential versions are considered more useful and practical for real world problems.
\\

\begin{figure}[h]
\centering
\includegraphics[width=0.95\textwidth]{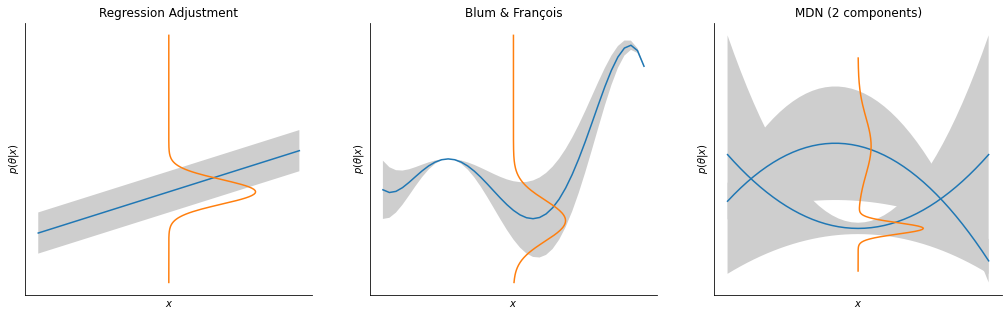}
\caption{Figure 1: Development of structure. The horizontal axis represents $x$, the vertical $\theta$. The blue line represents the mean of the posterior $p(\theta|x)$ and the gray shades represent the uncertainty / variance. The orange line is $p(\theta|x=1)$. Left: Regression Adjustment; Center: Blum \& François; Right: MDN with 2 components}
\end{figure}

Figure 1 depicts the structure discovery mechanism employed in SBI. The horizontal axis represents $x$, the vertical $\theta$. The blue line represents the mean of the posterior $p(\theta|x)$ and the gray shades represent the uncertainty / variance (say, 95\%). The orange line is $p(\theta|x=1)$. Moving from left to right depicts the development from linear regression and homoscedastic variance, to nonlinear regression with heteroscedastic noise, to a general structure that can have multiple modes, for example, a Mixture Density Network (MDN) where for every $x$, $p(\theta|x)$ is a Gaussian mixture model (GMM) with $k$ components ($k=2$ in the figure). \\

The sequential algorithms must correct for the fact that all rounds after the first one sample $\theta$ from proposal distributions that differ from the prior. SNPE-A\cite{snpe-a} restricted the prior to be a uniform or a Gaussian, and the NDE to be an MDN, and made a post-hoc correction to the Gaussians. In addition to these very limiting restrictions, the correction step sometimes failed to provide a valid distribution, causing the algorithm to crash. SNPE-B\cite{snpe-b} modified the loss function (negative log likelihood of the NDE) to include importance weights that adjust between the prior and the proposal. This had the great advantage of removing all the restrictions that were needed in SNPE-A; however, the importance weights were very dispersed, which increased the variance of the parameter updates during learning and led to slow or inaccurate inference. The approach in SNPE-C \cite{snpe-c} is to use a discrete approximation to the proposal and to sample, alongside each chosen $\theta$, several additional parameter vectors. This led them to a cross entropy loss, related to distinguishing the $\theta$ that generated $x_{sim}$ from the imposters. Figure 2 illustrates how the SNPE algorithm works. \\

\begin{figure}[h]
\centering
\includegraphics[width=0.9\textwidth]{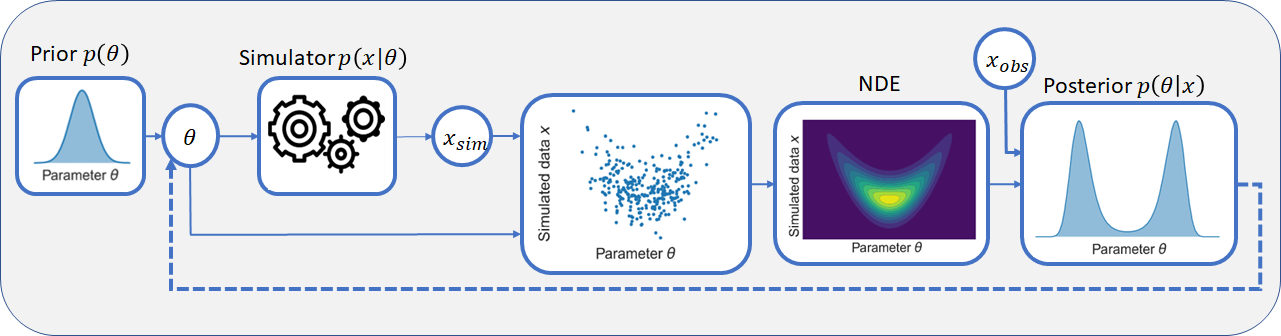}
\caption{Figure 2: Regular SNPE algorithm}
\end{figure}

Alternatively one can estimate or approximate the likelihood \cite{wood,fanetal,meeds} or the likelihood ratio  \cite{pham, cranmer}.  The initial approaches were called ``Synthetic Likelihood''. The SNLE method \cite{snle-a,snle-b} uses an NDE to model the likelihood $p(x|\theta)$ and employs sequential sampling.
It can be used on its own for frequentist inference or combined with a prior for Bayesian inference (e.g. by MCMC). Methods focused on the likelihood ratio use the fact that for computational Bayesian algorithms, such as Metropolis-Hastings MCMC, the likelihood is only used via ratios. These estimators exploit a powerful classifier to distinguish samples drawn from the joint distribution $p(x,\theta)=p(x|\theta)p(\theta)$ from samples drawn independently from the marginals $p(x)p(\theta)$. We can then use some basic algebraic manipulations to calculate the likelihood ratios we need as part of MCMC.  \\

These 3 algorithms - SNPE, SNLE, SNRE - (and their non-sequential counterparts), together with state of the art NDEs (e.g. normalizing flows such as Masked Autoregressive Flow (MAF) \cite{maf}, or Neural Spline Flows (NSF) \cite{nsf}) are considered the ``cutting edge'' algorithms in SBI. A recent benchmark paper \cite{benchmark} (and partially also \cite{durkan-early}) found that none of the algorithms is consistently superior to the others in terms of accuracy: different problems had different winners. \\

Two recent proposals use two NNs in estimating the simulator parameters: SNLPA \cite{SNPLA} and SNVI \cite{SNVI}. Both are similar in theory: in each round two NDE's are trained, one for the likelihood (or likelihood ratio) and one for the posterior. The likelihood NDE is optimized similarly to NLE/SNLE, by maximizing the log likelihood. The posterior NDE is updated in a VI fashion by optimizing the KL divergence between the posterior NDE $\hat p(\theta|x)$ and the likelihood NDE $\hat p(x|\theta)$ times the prior $p(\theta)$. \\

Our proposals incorporate ideas proposed in the related statistical research on the design and analysis of computer experiments \cite{dace}. Most of that research has
focused on deterministic simulators, for which a parameter vector $\theta$ always produces the same output $x(\theta)$. A recent review of the methods for stochastic simulators notes that ``there is a shortage of statistical research in this field'' (\cite{sto_sim}, p. 83). See Section \ref{NDE} for more detail. \\

\section{Reducing Simulator Calls} 

We examine here two different approaches with potential to reduce the calls to the simulator, thus accelerating SBI: \\

\begin{enumerate}
\item Use an NDE \textit{surrogate} $\hat p(x|\theta)$ instead of the simulator to generate $x(\theta)$;\\
\item Use \textit{support points} to better represent the distribution of $\theta$. \\
\end{enumerate}

\subsection{NDE Surrogate}\label{NDE}

A surrogate is an empirical approximation of the mapping from $\theta$ to $x$ in a simulator. Use of a cheap, and fast, surrogate has benefits when the simulator is complex and computationally expensive. See \cite{surrogates} for a broad presentation of the topic. Various algorithms can be used to produce surrogates: Gaussian Process (GP) regression, Neural Networks (NN), Trees, etc.
\\

Most of the research on surrogates has presented methods for deterministic simulators. Problems with stochastic simulators, for which
SBI is designed, present new challenges for using surrogates. In stochastic simulators, internal randomness maps each $\theta$ to a distribution of $x$'s and so requires surrogates that lead to a distribution. \\

Binois et al. \cite{hetGP_paper} developed an extension of GP surrogates for use with stochastic simulators. Their heteroscedastic GP method makes the simplifying assumption that the distributions $p(x|\theta)$ are all normal, or can be approximated well enough by a Gaussian. The big advantage is that the problem simplifies dramatically - the distributions are described completely by the mean $\mu(\theta)$ and the variance $\sigma^2(\theta)$. The method then proceeds by setting up separate GP regression models, one for estimating $\mu(\theta)$ and one for $\log(\sigma^2(\theta))$. The obvious limitation is that the resulting approximate likelihood may be inaccurate.\\

Stochastic heteroscedastic GP adds flexibility to homoscedastic GP by modeling Gaussians with different variances, but is still very limited: it assumes the distribution is uni-modal for every $\theta$; and it only allows for a single output. This is a problem in simulators that might exhibit multi-modal distributions, and have multi-dimensional outputs. It is possible to train a (heteroscedastic) GP regression for each output dimension, but this will result in a diagonal covariance Gaussian distribution: we lose the correlation across outputs, and we retain the uni-modality problem. There are also methods which exploit the output correlations, allowing the outputs to leverage information from one another. These are called Multiple Output GP (MOGP) (also known as multivariate Kriging or Co-Kriging) \cite{mogp}. These methods still exhibit the uni-modality problem. \\

Our novel proposal is to use Neural Density Estimators as surrogates for the stochastic simulator. This idea shares some common ground with NLE/SNLE, which use a single NDE to approximate the likelihood and then conduct MCMC to get an approximate posterior. Our method uses two NDE's: one for the likelihood, which acts as a surrogate for the simulator, and another for the posterior. In essence we are conducting SNPE, with the key difference that we replace the simulator with the surrogate after the 1st round:\\

\begin{itemize} 
\item In the 1st round of SNPE, feed the $\theta$'s into the simulator, yielding the data pairs $(\theta_i, x_{sim}(\theta_i))$;\\

\item Using the same data, train an NDE for the likelihood $\hat p(x|\theta)$, and another for the posterior $\hat p(\theta|x)$;\\

\item Continue with SNPE - in subsequent rounds sample $\theta$ from the proposal (the posterior after the last round), but rather than using $\theta$ as input to the simulator, feed it into the likelihood NDE $\hat p(x|\theta)$ and get a sample of $x_{sur}(\theta_i)$'s; \\

\item Use the new data $(\theta_i, x_{sur}(\theta_i))$ to retrain the posterior NDE, as in regular SNPE. Repeat the process for as many rounds as needed for the problem.\\
\end{itemize} 

This is the 
strategy we tested to distribute the simulator budget, but 
other strategies 
could also be adopted: e.g., spread it over multiple rounds in a ``warp and woof'' fashion, i.e., use the simulator on the 1st round to train the surrogate, use the surrogate on the 2nd round to sample points, use the simulator on the 3rd round to retrain the surrogate, use the surrogate on the 4th round to sample points, etc. \\

\subsection{Support Points} 

Support points \cite{sp} are a way to represent a distribution with a selected number of its ``most representative'' points. There are different algorithms to find representative points. Support Points belongs to the energy-representative-points class, as it optimizes the ``energy distance'' \cite{energy}. \\

Suppose we have a very large sample $y_{m=1:N}\in \mathbb R^d$, and want to replace it with a
small representative sample of size $n<N$. We do this by choosing a set of points $x_{i=1:n}$ that are 
as close as possible to the original points $y_{m=1:N}$
but as far apart from each other as possible:

$$\arg\min_{x_1,...,x_n} \frac{2}{nN}\sum_{i=1}^n \sum_{m=1}^N ||y_m-x_i||_2-\frac{1}{n^2}\sum_{i=1}^n\sum_{j=1}^n ||x_i-x_j||_2
$$

The criterion is a difference of convex functions, which allowed Mak and Joseph (\cite{sp}) to exploit efficient numerical algorithms. The key step in them is to replace the second term by a linear (Taylor series) approximation, resulting in a convex objective function. \\

We exploit SP SNPE as follows:\\

\begin{itemize} 
\item In each round of SNPE, draw a large sample $\theta_1,\ldots,\theta_N$ from the proposal $\tilde p(\theta)$ (either the prior, or last round posterior);\\

\item Pass the sampled points through the SP algorithm and reduce it to $xx\%$ most representative points of the original sample;\\

\item Feed only the SP points to the simulator, and continue regularly by training a posterior NDE on the data pairs.\\

\end{itemize}

\subsection{Combining a Surrogate with Support Points}
In order to test if there is a synergy from using both of the suggested methods together, we also assessed a method that combines them:\\
\begin{itemize}
    \item In the 1st round (or any real simulator round) use Support Points to get a sample of representative points. Feed that sample to the simulator, and train both a surrogate to replace the simulator and an NDE posterior to be the next round proposal distribution.\\
    \item In the 2nd round (or any surrogate round), sample from the proposal distribution and feed these to the surrogate to get $x_{sur}$. Use the additional data to train a new NDE posterior.\\
\end{itemize}

Figures 3-4 show the different algorithms visually: the Surrogate method (Figure 3), and the support-points method (Figure 4).

\begin{figure}[h]
\centering
\includegraphics[width=0.9\textwidth]{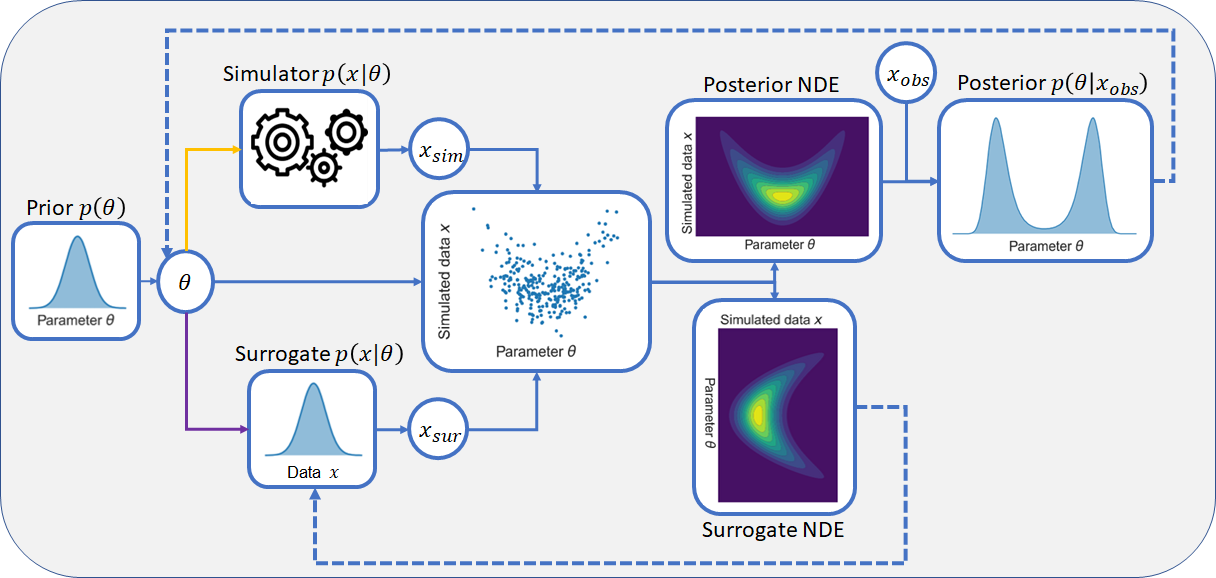}
\caption{Figure 3: Surrogate method - the orange path is used in simulator rounds, the purple path when the surrogate is used instead}
\end{figure}

\begin{figure}[h]
\centering
\includegraphics[width=0.9\textwidth]{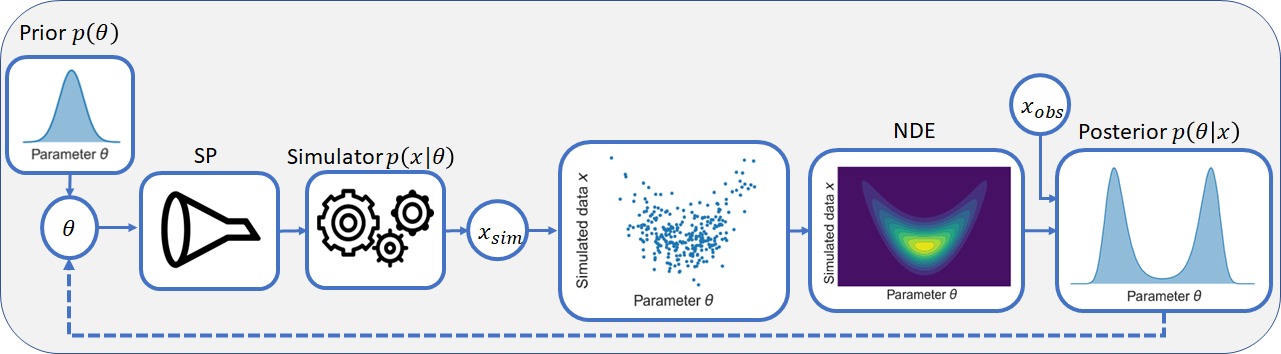}
\caption{Figure 4: Support Point Method on SNPE}
\end{figure}

\section{Test Cases}

We tested the new methods proposed in the last section (NDE surrogate, Support Points) on a set of six problems for which computing the actual posterior is possible. These tasks appeared in the related literature and five of them were also used in the benchmark paper \cite{benchmark}.  We compared the new methods to the regular method (SNPE-C) for \textbf{the same} simulator budgets. We used 3 different metrics that assess how close the result of the algorithm is to the true posterior.

\subsection{Tasks}
We used the following test problems to compare our new methods with the existing methods. Brief descriptions of each problem are given below; more details can be found in appendix A.\\

\begin{itemize}
    \item \textbf{Gaussian Mixture Model with same mean and different variance (1D)} \cite{snpe-a} - The true posterior is a 50\%-50\% GMM with mean $x$ and standard deviations of $0.1$ and $1$, truncated to the interval $[-10,10]$.\\
    
    \item \textbf{Bayesian Linear Regression (6D) with 10 observations} \cite{snpe-a} - Prior is a standard 6D multivariate normal $p(\theta) = \mathcal N(0,I)$. The covariates were sampled independently from the standard normal distribution 10 times. Finally $y|\theta\sim \mathcal N(y; X\theta, 0.1\cdot I)$. The posterior is also a multivariate Gaussian.\\ 
    
    \item \textbf{Two Moons (2D)} \cite{snpe-c} - A two-dimensional posterior that exhibits both global (bimodality) and local (crescent shape) structure. The prior is uniform on a square $\theta\sim U(-1,1)^2$. The simulated data are bivariate and concentrated near a random arc of radius $\sim 0.1$. The focal point of the arc depends on the parameters via a formula that leads to a bimodal posterior, with each mode crescent-shaped.\\

    \item \textbf{Simple Likelihood Complex Posterior (SLCP) (5D)}  \cite{snle-b} - The prior is uniform on a 5D box $\theta \sim U(-3,3)^5$. The data are 8 independent bivariate normal observations, whose means and covariance matrix are defined by the parameters. The construction is such that the posterior has four completely symmetric modes.\\ 
    
    \item \textbf{Bernoulli GLM (10D)} \cite{snpe-b} - Bernoulli observations are generated from a Gaussian white-noise series of size 100. The probability of response is given by a logistic regression model whose linear predictor includes an intercept and a window of size 9, moving through the series.\\
    
    \item \textbf{Sisson's $2^D$ GMM} \cite{sisson} - The data $x$ is a single $D$-variate observation drawn from a GMM with $2^D$ components; we used $D=3$. The parameter is also $D$-variate and determines the means of the components; the covariance matrix is assumed known. The prior is $\theta \sim U(-20,40)^D$.\\
\end{itemize}

Note that although increasing the dimensionality usually increases the complexity of the inference, there are also other factors at play. For example, Sisson's 3D task was usually more challenging for the algorithms to infer well, compared to the 6D Bayesian Linear Regression or the 10D Bernoulli GLM - probably due to the presence of multiple modes, instead of the single mode of the higher dimensional problems. 
\\

In addition we also tested the following:\\

\begin{itemize}
    \item \textbf{Compare the Surrogate method to SNLE} - we used both the surrogate method and the SNLE method on the 5D-SLCP problem. SNLE uses an NDE to approximate the likelihood and then continues by using MCMC to obtain samples from the posterior. We instead used the likelihood NDE as a surrogate and sampled more data points from it, before training a posterior NDE on the whole data. The SLCP problem was chosen as it is a problem whose likelihood structure is simpler and where SNLE is supposed to perform better. We wanted to compare it to the surrogate method. \\

    \item \textbf{Hodgkin-Huxley model (8D)} - this simulator model describes action-potential in neurons. The problem was first used in SBI in \cite{snpe-b} with 10 parameters, but later was presented in \cite{eLife} with 8 - that is the version that we implemented. The simulator creates an electrical trace of ``action potential''. The trace is then summarized by 13 statistics: the number of spikes; 5 auto-correlations (starting at 0.1-0.5 seconds from the action potential); the resting potential mean, its standard-deviation, and the mean current during action potential; and the 4 centralized moments (2nd, 3rd, 4th and 5th order) of the action potential trace (we used 13 dimensions instead of the 7 used in \cite{eLife}). The 8 parameters are unknown conductance terms; other conductance terms are presumed fixed/known. The prior distribution over the parameters is uniform and centered around the true parameter values: $\theta \sim U(0.5\theta^*, 1.5\theta^*)$. We used the same true parameters as in \cite{eLife}, namely: $(g_{Na}, g_K, g_l, g_M, T_{max}, -V_{T}, \sigma, -E_l) = (50, 5, 0.1, 0.07, 600, 60, 0.1, 70)$. We tested only the surrogate method, using both MDN and NSF. \\

    \end{itemize}

\subsection{Metrics}

In all of the constructed test problems we had access to the true posterior and were able to sample from it; thus we could compare the ``distance'' between the ground truth and the proposed new method (regular vs. NDE surrogate, regular vs. SP) to get an indication if the new method improves the inference. 
\\

Several metrics have frequently been 
used in the literature. These include the Total Variation (L1 distance), KL-Divergence, etc. These metrics usually require access to the full (i.e., analytical) distributions. In this work our access to the full analytical distributions is limited (both because some tasks only allowed us to get samples, but not compute the true posterior; and because the NDE's sometimes output the unnormalized estimated posterior and allow us only to sample from it). Consequently we preferred metrics that calculate distance between samples from distributions. 
\\

We used three metrics to assess accuracy. Two of them  (MMD, C2ST) were used by \cite{benchmark}.  We added a third metric (ED):\\
\begin{enumerate}
    \item \textbf{MMD - Maximum Mean Discrepancy} \cite{mmd}: represents the largest difference in expectation over functions in the unit ball of a Reproducing Kernel Hilbert Space (RKHS). Specifically we used 
    a Gaussian kernel, setting the bandwidth using the median heuristic\cite{ramdas}. Note we used the squared MMD. \\
    \item \textbf{C2ST - Classifier 2 Sample Test} \cite{c2st1,c2st2}: we train a (NN) classifier to distinguish whether or not the samples came from the true posterior distribution. The metric is the accuracy (fraction of correct classifications) with which the classifier distinguishes between the samples. The lowest accuracy (0.5) corresponds to the lowest possible distance between the samples, and high accuracy (close to 1.0) means the distance between the samples is high. One advantage of this metric is that it's absolute - a score close to 0.5 indicates a very good match between the true posterior sample and the test method sample. We used the same implementation as in \cite{benchmark}, a NN classifier (multi-layer-perceptron) using the Scikit Learn (sklearn) python library. \\
    \item \textbf{ED - Energy Distance} \cite{energy}: The squared energy distance between two samples $(\theta_1, \theta_2)$ (where, e.g., $\theta_1$ is a true posterior sample, and $\theta_2$ is a sample from an alternative method) is defined to be 
    $$ED^2(\theta_1,\theta_2) = 2\mathbb E[||\theta_1-\theta_2||] - \mathbb E[||\theta_1-\theta_1'||] - \mathbb E[||\theta_2-\theta_2'||].$$
    For computation, expectations are replaced by sample averages and the norm is the pair-wise distances between the points. That is, $||\theta_1-\theta_1'||$ is a vector of all the pair-wise distances between two points from the first sample, and $||\theta_1-\theta_2||$ is a vector of all the pair-wise distances of points when one is from each sample. The support points method generates a sub-sample by minimizing its ED from the original sample. Also note that the squared ED is very similar to the squared MMD and might be considered a special case of MMD: where we used the Gaussian kernel in the MMD, we apply a negative Euclidean distance kernel in the ED.\\
\end{enumerate}

For the Hodgkin-Huxley (HH) model, we do not have access to a true posterior. We proposed four reasonable metrics, two of which highlight closeness of location and two of which focus on dispersion. \\
\begin{enumerate}
    \item \textbf{M1 -  Distance between the true parameter and the sample median} - The true parameter $\theta_{true}$ refers to the parameter used to create $x_{obs}$. The median was calculated per dimension of the 8D vector. We used the L2 norm to calculate the distance. \\
    \item \textbf{M2 -  Distance between the true parameter and the sample mean} -  Similar to M1 only using the mean instead of the median. \\
    \item \textbf{M3 - Average of the Standard Deviations} - we averaged the standard deviations of the sample across the number of dimensions (8 in this case). \\
    \item \textbf{M4 - Average of the inter-quantile range} - we calculated the sample quantile values [0.15, 0.85]  for each dimension, found their difference, and then averaged over the 8 dimensions.\\
\end{enumerate}

As each parameter in the 8-dimensional posterior vector has a different range and variance, the L2 norms, SD, and IQR were scaled, dividing them by the corresponding prior range. 

\subsection{Simulator Budget}

We tested the new methods using different simulator budgets. How large a budget is needed to achieve good results varies from one problem to another. For example, 500 sample points achieved a very good C2ST score for the 1D GMM problem ($\approx$0.55), a moderate score for the 2D Two-Moons problem ($\approx$0.7), but
gave poor results for the 6D Bayesian Linear Regression problem ($\approx$0.9). As such we varied the sample sizes to match the problem complexity. Table 1 presents the budgets used for each problem.

\begin{table}[ht]
\centering
\small
\begin{tabular}{|l|l|}
\hline
\textbf{Problem Name}       & \textbf{Budgets}                                                                                   \\ \hline
1D-GMM                      & 100, 200, 300, 400, 500, 750, 1000, 1250, 1500, 1750, 2000                                         \\ \hline
2D-Two-Moons                & 100, 200, 300, 400, 500, 750, 1000, 1250, 1500, 1750, 2000                                         \\ \hline
3D-Sisson                   & 250, 500, 1000, 1500, 2500, 5000, 7500                                                             \\ \hline
5D-SLCP                     & 250, 500, 1000, 2500, 5000, 7500                                                                   \\ \hline
6D-Bayesian-LR              & 250, 500, 750, 1000, 1500, 2500, 5000, 7500, 10000                                                 \\ \hline
10D-Bernoulli-GLM           & 250, 500, 1000, 1500, 2500, 5000, 7500, 10000                                                      \\ \hline
SNLE vs. Surrogate          & 250, 500, 1000, 2500, 5000, 7500, 10000                                                            \\ \hline
HH                          & 1000, 2500, 5000, 15000, 25000                                                                     \\ \hline
\end{tabular}
\caption{Table 1: Budgets tested for various problems.}
\label{table:problem_budgets}
\end{table}

Due to time and resource constraints, we used a limited budget on the HH problem. It should be noted that the highest budget used in this paper (25,000, divided into 2 rounds of SNPE) is still significantly lower than the one used in \cite{eLife} (100,000, used in a single round). 
\\

The basic premise in this paper is that calls to the simulator in real world complex applications are much more expensive than sampling from a surrogate. As such we can generate many samples from the surrogate without needing to increase the simulator budget. In our experiments the sample from the trained surrogate was 10 times larger than the simulator sample.\\

\subsection{Initial Exploration}

For the surrogate method, we tested spending all the simulation budget on the first round to a two-round process which used the surrogate to sample more points. As this initial exploration showed promising results, we continued with it in the full experiments.
\\

For the support points method, we compared initial samples of size $2n, 5n, 10n, 20n$ on the 1D GMM and 2D Two Moons problems. Although the support points generally were better than the regular method, improvement was not monotonic, so we opted to use $2n$ for the initial sample. \\

In addition to using a surrogate NDE, we also tried other surrogate methods, including deterministic function approximators (such as a feed-forward NN, implemented using PyTorch), and a stochastic approximator. The latter applied the approach in \cite{hetGP_paper}, which involves fitting a heteroscedastic Gaussian Process to the $x|\theta$ data and then sampling $x$ from the relevant Gaussian; the implementation was done using the python `rpy2' library which interfaced with the R `hetGP' library \cite{hetGP}). The results here from limited experimentation were worse than using a full-fledged NDE, which has more flexibility for generating output 
distributions whereas `hetGP' is forced to be  
a Gaussian, and the deterministic NN can only output a single number resulting in extreme under-dispersion of the posterior. Note that we ran tests only on the 1D GMM problem because the `hetGP' library is limited to a 1D output. We could circumvent this limitation by running `hetGP' independently for each output; however, that would result in a diagonal-covariance Gaussian, which is obviously much less flexible than an NDE. 
Even with a single output, our results were less successful, so we did not pursue 
comparison to this method.
\\

\section{Results}

\subsection{Setup}
The experiments were conducted on Google Cloud platform with C2D machines (2 vCPU (1 core), 8 GB memory). Code was written in Python, with use of PyTorch and the SBI libraries, and is publicly available in GitHub \cite{code}. For the HH problem, the simulator was run using the \cite{eLife} Cython implementation, which was orders of magnitude faster than a Python implementation. For the toy tasks, we ran 5-20 experiments per problem-budget pair (quicker problems allowed for more replicated experiments, while slower problems allowed for less). For the SNLE vs. Surrogate problem we used 10 experiments. For the HH simulator we used 6-10 experiments.\\

The SBI algorithms are very modular and there are many moving parts. It would be very difficult to try out all the possible configurations. A limited initial exploration led us to choose the sequential algorithms as they are more realistic for complex real world problems. Also, since no significant differences were found in the exploration between the SNPE, SNLE and SNRE algorithms (matching the report in \cite{benchmark}), we opted to use SNPE, which is generally faster for modelling the posterior directly. In particular, we used SNPE-C, the most mature algorithm, which is the default in the SBI library. \\

Based on initial exploration of the 3 available NDE's in the SBI library (at the time): MDN, MAF and NSF, we chose NSF, which showed the best overall initial performance. \\

The task experiments compared the regular SNPE method to the surrogate method, to the SP method, and to 
the combined surrogate and SP.
In the tasks, all experiments were run with 2-rounds of SNPE-C.\\

The SNLE experiments ran 2 rounds of either SNLE or the new surrogate method on the 5D-SLCP problem. Two types of NDE were tested: MDN, which is faster to sample from, but less expressive, and NSF, which is more expressive but slower to sample from. \\

The HH experiments ran 2 rounds of either SNPE or the surrogate method. Here too we compared both using MDNs and NSFs. \\

\subsection{Metrics Reduction}

The methods were tested multiple times, and the mean results were calculated for the 3 chosen metrics, in addition to the standard deviation (SD), and the improvement ratio (number of runs where the new method performed better versus number of runs where the regular method was better). As such 9 final metrics were given per problem-budget scheme. If at least 5 showed improvement we considered the overall result to be better. If 4 or less showed improvement we consider the result worse. \\

The 3 metrics (MMD, C2ST and ED) are positively correlated (especially MMD and ED) for both mean and median scores, and exhibit moderately high correlation for the individual runs. For each task, correlations for each pair of metrics
were computed for each budget and each analysis method, and then were averaged within task. The average correlations ranged from 0.46 for the 3D-Sisson problem, to 0.82 for the 6D Bayesian LR. \\

Increasing the simulator budgets had a consistent effect: adding more sample points improved the results for both the regular method and the new methods. The notable exception was the 6D Bayesian Linear Regression, which showed some non-monotonic behavior, especially for the regular scheme. \\

\subsection{NDE Surrogate}

To code up the NDE surrogate, we simply used the NSF implementation of the SBI library, reversing the roles of $x$ and $\theta$ (feeding $x$ as $\theta$ and $\theta$ as $x$). This resulted in a conditional NDE of the form $p(x|\theta)$.
\\

The experiments were conducted with the SNPE-C algorithm, with 2 sequential rounds. That is, for the regular method we run 2 rounds of SNPE: in the 1st round we train an NDE posterior, and then use it in the 2nd round as the proposal from which to draw $\theta$. The simulator budget is thus divided between the 2 rounds. In the new proposed method, in the 1st round we train 2 NDE's, one to act as a surrogate for the simulator (in which $\theta$ and $x$ switch roles) and one for the regular posterior NDE. Here the simulator budget is spent completely in the 1st round. In the 2nd round we sample from the $\theta$ proposal, and then conditional on the $\theta$, we sample $x$ from the NDE surrogate, and add the data to train the posterior NDE. \\

The results are summarized below for each task. The summary begins by computing summaries for each budget -- the number of metrics (out of 9) in which the proposal improved performance, and for C2ST, the average and median improvement. The C2ST averages and medians are then summarized over budgets by average and range.  For most of the test problems, NDE had better results than the standard method for some budgets but worse performance for others. In all of these cases, the improvements were much more dramatic than the drops in performance.\\

\begin{itemize}
    \item 1D GMM - Surrogate NDE showed good results (6/9 and greater) for 7 out of 11 budgets: 100, 200, 300, 400, 500, 1500, 2000 and bad results (4/9 and less) for 750, 1000, 1750, 1250. The C2ST for the regular SNPE method score improved from $0.71$ for a budget of 100, to $0.52$ for a budget of 2000. 
    For the averages, the surrogate method had results ranging from $0.015$ worse, to $0.027$ better than the regular method, with an average improvement of $0.004$. The matching results for the medians: from $0.014$ worse to $0.041$ better, with an average improvement of 0.005. \\

    \item 2D Two Moons - Surrogate NDE showed improvements (5/9 and greater) for 8 out of 11 budgets: 200, 300, 400, 500, 750, 1000, 1500, 2000 and bad results (4/9 and 3/9) for 100, 1250, 1750. The medians and IQRs showed good results for 9 out of 11: changing the 100 and 1250 budgets to good results (5/9) and 1500 budget to a bad result (4/9). The C2ST for regular SNPE decreased from $0.92$ for a budget of 100 to $0.59$ for a budget of 2000. The corresponding results for the medians were 13.45\% (3.1\%). For the averages, the surrogate method had results ranging from $0.027$ worse, to $0.135$ better than the regular method, with an average improvement of $0.035$. The matching results for the medians: from $0.014$ worse to $0.041$ better, with an average improvement of 0.032.\\

    \item 3D Sisson's $2^D$ GMM - Surrogate NDE showed improvements (5/9 and greater) for 6 out of 7 budgets: 250, 500, 1000, 2500, 5000, 7500 and bad results (1/9) for 1500. The medians and IQRs were a bit worse: 5 out of 7 budgets were good, the 2500 budget became a bad result (1/9). The C2ST for regular SNPE decreased from $0.94$ for a budget of 250 to $0.7$ for a budget of 7500.     For the averages, the surrogate method had results ranging from $0.038$ worse, to $0.021$ better than the regular method, with an average loss of $0.004$. The matching results for the medians: from $0.038$ worse to $0.014$ better, with an average loss of 0.006.
    \\

    \item 5D SLCP - Surrogate NDE showed improvements (6/9 and greater) for 4 out of 6 budgets: 1000, 2500, 5000, 7500 and bad results (2/9 and 3/9) for 250 and 500. The number of bad rounds for the 250 and 500 budgets were significantly higher than good rounds, making the ratio of good rounds across all budgets 0.48, 0.56 and 0.47 for MMD, C2ST and ED respectively. The C2ST for regular SNPE decreased from $0.95$ for a budget of 250 to $0.82$ for a budget of 7500. For the averages, the surrogate method had results ranging from $0.002$ worse, to $0.033$ better than the regular method, with an average improvement of $0.011$. The matching results for the medians: from $0.002$ worse to $0.026$ better, with an average improvement of 0.009.\\

    \item 6D Bayesian Linear Regression - Surrogate NDE showed improvements (6/9 and greater) for 8 out of 9 budgets: 250, 500, 750, 1000, 1500, 2500, 5000, 10000, and bad results (3/9) for 7500. The medians and IQRs show improvements in only 6 out of 9 budgets: the 2500 and 5000 budgets became bad (2/9). The C2ST for regular SNPE decreased from $0.99$ for a budget of 250 to $0.75$ for a budget of 10000. For the averages, the surrogate method had results ranging from $0.018$ worse, to $0.091$ better than the regular method, with an average improvement of $0.032$. The matching results for the medians: from $0.044$ worse to $0.069$ better, with an average improvement of 0.018.
    \\
    
    \item 10D Bernoulli GLM - 
    only 3 out of the 8 budgets were good (5/9 and above): 250, 1000, 1500. The rest of the budgets were bad (500, 2500, 5000, 7500, 10000), even 
    reaching 0/9. The C2ST for regular SNPE score decreased from $0.95$ for a budget of 250 to $0.82$ for a budget of 7500.   For the averages, the surrogate method had results ranging from $0.033$ worse, to $0.044$ better than the regular method, with an average improvement of $0.003$. The matching results for the medians: from $0.078$ worse to $0.030$ better, with an average loss of 0.009.
    \\

\end{itemize}

Table 2 shows the average reduction to the means and SDs, and average Good-vs.-Bad rounds ratio, taken across the different budgets used per problem for the NDE surrogate method. Table 3 shows the same for the medians and IQRs. It is worth contrasting the differences here to the ones presented in \cite{benchmark}. In \cite{benchmark}, for most problems the differences in C2ST (across different algorithms: SNPE, SNLE, SNRE, etc.) seem to be $0.05-0.1$ (except for the Bernoulli GLM task which showed $\approx 0.2$ difference) - see figure 2 and figure 3 in \cite{benchmark}. Our methods show milder differences in the C2ST metric (mean difference is $0.015$, median is $0.01$, maximal difference was $0.12$ for means and $0.1345$ for the medians).

\begin{table}[ht]
\centering
\small 
\begin{tabular}{|l|ccc|ccc|ccc|}
\hline
\textbf{Problem} & \multicolumn{3}{c|}{\textbf{Mean Reduction}} & \multicolumn{3}{c|}{\textbf{SD Reduction}} & \multicolumn{3}{c|}{\textbf{Good:Bad Ratio}} \\ \hline
 & \textbf{MMD} & \textbf{C2ST} & \textbf{ED} & \textbf{MMD} & \textbf{C2ST} & \textbf{ED} & \textbf{MMD} & \textbf{C2ST} & \textbf{ED} \\ \hline
1D GMM & 0.0038 & 0.0036 & 0.0151 & 0.0042 & {-0.0002} & 0.0256 & 0.5636 & 0.5636 & 0.5364 \\ \hline
2 Moons & 0.0037 & 0.0354 & 0.00273 & 0.00246 & 0.0103 & 0.00091 & 0.5636 & 0.6636 & 0.5318 \\ \hline
3D Sisson & 0.00533 & {-0.00435} & 0.11073 & 0.00073 & 0.00822 & 0.02690 & 0.62044 & 0.44526 & 0.60584 \\ \hline
5D SLCP & 0.00622 & 0.01113 & 0.00583 & 0.00637 & {-0.00639} & 0.03355 & 0.48571 & 0.56190 & 0.47619 \\ \hline
6D Bayes LR & 0.06436 & 0.03233 & 0.12218 & 0.09004 & 0.02805 & 0.17914 & 0.64368 & 0.65517 & 0.58621 \\ \hline
10D BerGLM & {-0.00279} & 0.00311 & {-0.02984} & 0.00174 & 0.00397 & {-0.03878} & 0.37500 & 0.50000 & 0.38750 \\ \hline
\end{tabular}
\caption{Table 2: Averaged Across Different Budgets for various problems (Mean Reduction, SD Reduction, Good:Bad Ratio).}
\label{table:averaged_across_budgets}
\end{table}

\begin{table}[ht]
\centering
\small 
\begin{tabular}{|l|ccc|ccc|ccc|}
\hline
\textbf{Problem} & \multicolumn{3}{c|}{\textbf{Median Reduction}} & \multicolumn{3}{c|}{\textbf{IQR Reduction}} & \multicolumn{3}{c|}{\textbf{Good:Bad Ratio}} \\ \hline
 & \textbf{MMD} & \textbf{C2ST} & \textbf{ED} & \textbf{MMD} & \textbf{C2ST} & \textbf{ED} & \textbf{MMD} & \textbf{C2ST} & \textbf{ED} \\ \hline
1D GMM & 0.0033 & 0.0048 & 0.0106 & -0.0032 & -0.0057 & -0.0040 & 0.5636 & 0.5636 & 0.5364 \\ \hline
2 Moons & 0.00193 & 0.03177 & 0.00236 & -0.00039 & 0.01286 & -0.00150 & 0.5636 & 0.6636 & 0.5318 \\ \hline
3D Sisson & 0.00593 & -0.00614 & 0.11917 & 0.00136 & 0.01146 & 0.03698 & 0.62044 & 0.44526 & 0.60584 \\ \hline
5D SLCP & 0.00766 & 0.00917 & -0.00651 & 0.00205 & -0.01840 & 0.00678 & 0.48571 & 0.56190 & 0.47619 \\ \hline
6D Bayes LR & 0.01916 & 0.01811 & 0.04526 & 0.04320 & 0.01433 & 0.04088 & 0.64368 & 0.65517 & 0.58621 \\ \hline
10D BerGLM & -0.00829 & -0.00856 & -0.03465 & 0.00552 & 0.00377 & 0.01233 & 0.37500 & 0.50000 & 0.38750 \\ \hline
\end{tabular}
\caption{Table 3: Averaged Across Different Budgets for various problems (Median Reduction, IQR Reduction, Good:Bad Ratio).}
\label{table:averaged_across_budgets_median}
\end{table}

\subsection{Support Points}

We coded the Support Points (convex-concave procedure - CCP) algorithm manually. Although this algorithm can be parallelized, we implemented a non-parallel version in python.\\

When used as part of SBI, the Support Points method samples $2n$ points from the proposal, and then uses the Support Points algorithm to reduce this sample to the most representative $n$ points. We used a 2-round SNPE-C algorithm with the NSF as the choice for the NDE.\\

The results are as follows (the C2ST for the regular SNPE method scores are as stated above):\\

\begin{itemize}
    \item 1D GMM - SP showed good results (5/9 and greater) for 9 out of 11 budgets: 100, 300, 400, 500, 750, 1000, 1500, 1750, 2000, and bad results for 200, 1250. The improvement in the good budgets was greater than the harm to performance in the bad budget in all metrics. Looking at medians and IQR's worsens the situation, though only turning the 1000 budget into a bad result (4/9).     For the averages, the support points had results ranging from $0.012$ worse, to $0.016$ better than the regular method, with an average improvement of $0.005$. The matching results for the medians: from $0.021$ worse to $0.034$ better, with an average improvement of 0.009.\\

    \item 2D Two Moons - SP showed improvements (5/9 and greater) for only 4 out of 11 budgets: 300, 500, 1500, 2000 and bad results (up to 0/9) for 100, 200, 400, 750, 1000, 1250, 1750. The medians and IQRs show improvement in the 400 budget turning it to a good result. The results were very disappointing here.       For the averages, the support points had results ranging from $0.028$ worse, to $0.042$ better than the regular method, with an average loss of $0.004$. The matching results for the medians: from $0.036$ worse to $0.053$ better, with an average loss of 0.008.\\

    \item 3D Sisson's $2^D$ GMM - SP showed improvements (5/9 and greater) for 5 out of 7 budgets: 500, 1500, 2500, 5000, 7500 and bad results (3/9 and 1/9) for 250 and 1000. Improvement was less dramatic than the loss in performance for bad budgets. In terms of the medians and IQRs, the 250 budget becomes a good result, but the 5000 budget becomes a bad result.      For the averages, the support points had results ranging from $0.020$ worse, to $0.013$ better than the regular method, with an average loss of $0.004$. The matching results for the medians: from $0.013$ worse to $0.004$ better, with an average loss of $0.006$.\\

    \item 5D SLCP - SP showed improvements (9/9 and 5/9) for the budgets 500 and 5000. The rest of the budgets showed bad results (up to 0/9): 250, 1000, 2500, 7500. Medians and IQRs show improvement in the 7500 budget turning it to a good result (6/9). On this problem also the results were disappointing.     For the averages, the support points had results ranging from $0.002$ worse, to $0.037$ better than the regular method, with an average improvement of $0.006$. The matching results for the medians: from $0.004$ worse to $0.074$ better, with an average improvement of $0.012$.\\

    \item 6D Bayesian Linear Regression - SP showed perfect results (9/9) for 4 out of 9 budgets: 1000, 1500, 2500, 10000, and bad results (up to 1/9) for 250, 500, 750, 5000, 7500. For all metrics, improvement was greater than the drop in performance. In terms of the medians and IQRs, the 500 budget was also a good result (5/9).      For the averages, the support points had results ranging from $0.029$ worse, to $0.077$ better than the regular method, with an average improvement of $0.015$. The matching results for the medians: from $0.041$ worse to $0.085$ better, with an average improvement of $0.007$.\\
    
    \item 10D Bernoulli GLM - SP showed good results (5/9 and 7/9) for 1000 and 5000 budget, but bad results (up to 0/9) for the rest of the budgets: 250, 500, 1500, 2500, 7500, 10000. Looking at medians and IQRs turns the 1000 budget into a bad result but the 7500 budget into a good result.      For the averages, the support points had results ranging from $0.012$ worse, to $0.017$ better than the regular method, with both methods equal on average. The matching results for the medians: from $0.026$ worse to $0.016$ better, with an average loss of 0.002.\\
\end{itemize}

Table 4 shows the average reduction to the means and SDs, and average Good-vs.-Bad rounds ratio, taken across the different budgets used per problem for the SP method. Table 5 shows the same for medians and IQRs. \\

\begin{table}[ht]
\centering
\small 
\begin{tabular}{|l|ccc|ccc|ccc|}
\hline
\textbf{Problem} & \multicolumn{3}{c|}{\textbf{Mean Reduction}} & \multicolumn{3}{c|}{\textbf{SD Reduction}} & \multicolumn{3}{c|}{\textbf{Good:Bad Ratio}} \\ \hline
 & \textbf{MMD} & \textbf{C2ST} & \textbf{ED} & \textbf{MMD} & \textbf{C2ST} & \textbf{ED} & \textbf{MMD} & \textbf{C2ST} & \textbf{ED} \\ \hline
1D GMM & 0.0029 & 0.0052 & 0.0129 & 0.0064 & 0.002359 & 0.0316 & 0.5182 & 0.5318 & 0.5318 \\ \hline
2 Moons & -0.00151 & -0.00455 & -0.00126 & -0.00158 & 0.00036 & -0.00141 & 0.4545 & 0.4455 & 0.4500 \\ \hline
3D Sisson & 0.00077 & -0.00374 & 0.01390 & -0.00061 & 0.00103 & -0.01199 & 0.55474 & 0.45985 & 0.53285 \\ \hline
5D SLCP & -0.00335 & 0.00617 & -0.03350 & -0.00729 & -0.00482 & -0.04404 & 0.47619 & 0.49524 & 0.45714 \\ \hline
6D Bayes LR & 0.01602 & 0.01522 & 0.02900 & 0.00613 & 0.01118 & 0.00932 & 0.54032 & 0.54032 & 0.57471 \\ \hline
10D BerGLM & -0.00836 & -0.00244 & -0.04252 & -0.01290 & -0.00138 & -0.06529 & 0.42500 & 0.43750 & 0.42500 \\ \hline
\end{tabular}
\caption{Table 4: Averaged Across Different Budgets for various problems (Mean Reduction, SD Reduction, Good:Bad Ratio).}
\label{table:averaged_across_budgets_sp_mean}
\end{table}

\begin{table}[ht]
\centering
\small 
\begin{tabular}{|l|ccc|ccc|ccc|}
\hline
\textbf{Problem} & \multicolumn{3}{c|}{\textbf{Median Reduction}} & \multicolumn{3}{c|}{\textbf{IQR Reduction}} & \multicolumn{3}{c|}{\textbf{Good:Bad Ratio}} \\ \hline
 & \textbf{MMD} & \textbf{C2ST} & \textbf{ED} & \textbf{MMD} & \textbf{C2ST} & \textbf{ED} & \textbf{MMD} & \textbf{C2ST} & \textbf{ED} \\ \hline
1D GMM & 0.0003 & 0.0085 & 0.0032 & 0.0020 & 0.004682 & 0.0085 & 0.5182 & 0.5318 & 0.5318 \\ \hline
2 Moons & -0.00228 & -0.00800 & -0.00216 & 0.00024 & -0.00164 & -0.00137 & 0.4545 & 0.4455 & 0.4500 \\ \hline
3D Sisson & 0.00179 & -0.00571 & 0.02690 & 0.00226 & 0.00227 & 0.02097 & 0.55474 & 0.45985 & 0.53285 \\ \hline
5D SLCP & 0.00442 & 0.01183 & 0.00347 & -0.02045 & -0.01304 & -0.10960 & 0.47619 & 0.49524 & 0.45714 \\ \hline
6D Bayes LR & 0.00148 & 0.00717 & 0.00331 & 0.00519 & 0.00124 & 0.00383 & 0.54032 & 0.54032 & 0.57471 \\ \hline
10D BerGLM & -0.00403 & -0.00212 & -0.01675 & -0.00123 & 0.00131 & -0.00911 & 0.42500 & 0.43750 & 0.42500 \\ \hline
\end{tabular}
\caption{Table 5: Averaged Across Different Budgets for various problems (Median Reduction, IQR Reduction, Good:Bad Ratio).}
\label{table:averaged_across_budgets_sp_median}
\end{table}

\subsection{Combined Algorithm}

The combined algorithm uses both support points and the surrogate method. The experiments were conducted using 2-rounds SNPE-C with NSF as the choice for the NDE. Since the surrogate method takes $10$ times as many samples from the surrogate, combining it with support points means we need to take $20$ times as many samples and then subsample with SP to reduce the size by a factor of $2$. This proved reasonable for the two small problems (1D GMM and  2D-Two-Moons), but it was very slow for the other problems. As such we decided to use the support points only in the 1st round where we take $2n$ samples and then subsample to $n$ - and train both NDE's (for the surrogate, and for the posterior/proposal). In the 2nd round we used the surrogate to sample $10n$, and didn't use SP. \\

The results are as follows (the C2ST for the regular SNPE method scores are as stated above). As with the surrogate alone, improvements in performance were more dramatic than were decreases.\\

\begin{itemize}
    \item 1D GMM - combined algorithm showed good results (5/9 and greater) for 7 out of 11 budgets: 100, 200, 400, 500, 750, 1500, 2000, and bad results for 300, 1000, 1250, 1750. 
    For the averages, the combined algorithm had results ranging from $0.019$ worse, to $0.043$ better than the regular method, with an average improvement of $0.015$. The matching results for the medians: from $0.019$ worse to $0.059$ better, with an average improvement of $0.004$.\\
    
    \item 2D Two Moons - combined algorithm showed improvements (5/9 and greater) for 9 out of 11 budgets: 200, 300, 400, 500, 750, 1000, 1250, 1500, 2000 and bad results (0/9 and 3/9) for 100 and 1750 respectively. Results were less favorable when comparing medians and IQRs, with worse results also for 750, 1000, and 2000.     For the averages, the combined algorithm had results ranging from $0.01$ worse, to $0.12$ better than the regular method, with an average improvement of $0.035$. The matching results for the medians: from $0.017$ worse to $0.1$ better, with an average improvement of $0.026$.\\

    \item 3D Sisson's $2^D$ GMM - combined algorithm showed improvements (6/9 and greater) for 5 out of 7 budgets: 250, 500, 2500, 5000, 7500 and bad results (4/9 and 3/9) for 1000 and 1500. In terms of the medians and IQRs, the 1000 budget was a good result, but the 2500 budget was a bad result.     For the averages, the combined algorithm had results ranging from $0.036$ worse, to $0.027$ better than the regular method, with an average loss of $0.003$. The matching results for the medians: from $0.039$ worse to $0.011$ better, with an average improvement of $0.007$.\\

    \item 5D SLCP - combined algorithm showed improvements (5/9 and greater) for only 3 out of the 6 budgets: 2500, 5000, 7500. The rest of the budgets showed bad results (up to 1/9): 250, 500, 1000. With medians and IQRs, the 2500 budget became a bad result (4/9). On this test case improvements were more or less equal to performance loss (a bit lower in the means/SDs, a bit higher in the medians/IQRs), but the ratio of results clearly favored the regular method here.     For the averages, the combined algorithm had results ranging from $0.014$ worse, to $0.022$ better than the regular method, with an average improvement of $0.0013$. The matching results for the medians: from $0.015$ worse to $0.029$ better, with an average improvement of $0.0036$.\\

    \item 6D Bayesian Linear Regression - combined algorithm showed good results (6/9 or greater) for 8 out of 9 budgets: 250, 500, 750, 1000, 1500, 2500, 5000, 10000 and bad results (3/9) for the 7500. Viewing medians and IQRs, the 2500 and 5000 were also bad results. Still improvements are clearly more dramatic.     For the averages, the combined algorithm had results ranging from $0.005$ worse, to $0.115$ better than the regular method, with an average improvement of $0.043$. The matching results for the medians: from $0.03$ worse to $0.1$ better, with an average improvement of $0.028$.\\
    
    \item 10D Bernoulli GLM - combined algorithm showed good results (5/9 to 9/9) for 4 out of 8 budgets: 250, 500, 1000, 1500. Bad results (3/9 to 0/9) were found for 2500, 5000, 7500 and 10000. In terms of medians and IQRs, the 500 budget was a bad result, but the 2500 budget was a good result. The improvements were greater than the performance drops, but the good:bad ratio clearly favors the regular method.     For the averages, the combined algorithm had results ranging from $0.03$ worse, to $0.054$ better than the regular method, with an average improvement of $0.006$. The matching results for the medians: from $0.039$ worse to $0.062$ better, with an average improvement of $0.0001$.\\

\end{itemize}

Table 6 shows a table of the average reduction to the means and SDs, and average Good-vs.-Bad rounds ratio, taken across the different budgets used per problem for the mix method. Table 7 shows the same for medians and IQRs. \\

\begin{table}[ht]
\centering
\small 
\begin{tabular}{|l|ccc|ccc|ccc|}
\hline
\textbf{Problem} & \multicolumn{3}{c|}{\textbf{Mean Reduction}} & \multicolumn{3}{c|}{\textbf{SD Reduction}} & \multicolumn{3}{c|}{\textbf{Good:Bad Ratio}} \\ \hline
 & \textbf{MMD} & \textbf{C2ST} & \textbf{ED} & \textbf{MMD} & \textbf{C2ST} & \textbf{ED} & \textbf{MMD} & \textbf{C2ST} & \textbf{ED} \\ \hline
1D GMM & 0.0030 & 0.0014 & 0.0119 & 0.0035 & 0.0026 & 0.0135 & 0.5182 & 0.5000 & 0.5273 \\ \hline
2 Moons & 0.0022 & 0.0349 & 0.0019 & 0.0029 & 0.0056 & 0.0005 & 0.5591 & 0.6182 & 0.5364 \\ \hline
3D Sisson & 0.0059 & -0.0036 & 0.1231 & 0.0013 & 0.0057 & 0.0368 & 0.6058 & 0.4453 & 0.5912 \\ \hline
5D SLCP & 0.0002 & 0.0013 & -0.0327 & 0.0048 & -0.0008 & 0.0238 & 0.4286 & 0.4000 & 0.4190 \\ \hline
6D Bayes LR & 0.0794 & 0.0431 & 0.1455 & 0.1147 & 0.0306 & 0.2243 & 0.5862 & 0.6782 & 0.6092 \\ \hline
10D BerGLM & 0.0088 & 0.0066 & 0.0257 & 0.0204 & 0.0096 & 0.0805 & 0.3625 & 0.4750 & 0.3750 \\ \hline
\end{tabular}
\caption{Table 6: Averaged Across Different Budgets for various problems (Mean Reduction, SD Reduction, Good:Bad Ratio).}
\label{table:averaged_across_budgets_mix_mean}
\end{table}

\begin{table}[ht]
\centering
\small 
\begin{tabular}{|l|ccc|ccc|ccc|}
\hline
\textbf{Problem} & \multicolumn{3}{c|}{\textbf{Median Reduction}} & \multicolumn{3}{c|}{\textbf{IQR Reduction}} & \multicolumn{3}{c|}{\textbf{Good:Bad Ratio}} \\ \hline
 & \textbf{MMD} & \textbf{C2ST} & \textbf{ED} & \textbf{MMD} & \textbf{C2ST} & \textbf{ED} & \textbf{MMD} & \textbf{C2ST} & \textbf{ED} \\ \hline
1D GMM & 0.0008 & 0.0047 & 0.0079 & 0.0014 & 0.0016 & 0.0076 & 0.5182 & 0.5000 & 0.5273 \\ \hline
2 Moons & -0.0002 & 0.0265 & -0.0002 & 0.0016 & 0.0056 & -0.0003 & 0.5591 & 0.6182 & 0.5364 \\ \hline
3D Sisson & 0.0055 & -0.0076 & 0.1183 & 0.0031 & 0.0096 & 0.0612 & 0.6058 & 0.4453 & 0.5912 \\ \hline
5D SLCP & 0.0039 & 0.0036 & -0.0122 & 0.0033 & 0.0038 & 0.0125 & 0.4286 & 0.4000 & 0.4190 \\ \hline
6D Bayes LR & 0.0253 & 0.0286 & 0.0500 & 0.0576 & 0.0157 & 0.0539 & 0.5862 & 0.6782 & 0.6092 \\ \hline
10D BerGLM & -0.0012 & 0.0001 & -0.0178 & 0.0145 & 0.0134 & 0.0417 & 0.3625 & 0.4750 & 0.3750 \\ \hline
\end{tabular}
\caption{Table 7: Averaged Across Different Budgets for various problems (Median Reduction, IQR Reduction, Good:Bad Ratio).}
\label{table:averaged_across_budgets_mix_median}
\end{table}

\subsection{SNLE vs. Surrogate}

We compared the SNLE method to the surrogate method on one select task: 5D SLCP, which favors the SNLE method due to a very simple likelihood structure copmared to the posterior. The surrogate performed rather poorly when compared to the SNLE method. For full details check appendix B. 

\subsection{Hodgkin-Huxley}

We measured the 4 metrics specified above for 5 different budgets: 1000, 2500, 5000, 15000, 25000. All experiments were repeated 10 times, except for the 25000 budget which was only repeated 6 times due to resource and time constraints. \\

We have 4 metrics that we devised: 2 measure location, 2 measure dispersion. As the experiments were repeated, we have means and SDs of these measures. In addition there are 4 good:bad ratios for each metric, so overall there are 12 metrics for each budget in the HH problem.  If 6 or more were good, we consider it a good result, if 5 or less - we consider it a bad result.\\ 

\begin{itemize}
    \item \textbf{MDN} - 2500, 5000 and 25000 were good results. 1000 and 15000 were bad. Improvements were much more dramatic than performance drops, and the good:bad ratio favored the surrogate method for all but the M4 metric.  \\
    \item \textbf{NSF} - 15000 and 25000 had very good results. 1000 and 5000 had borderline (6/12), and 2500 had bad results (4/12). \\

\end{itemize}

Surprisingly the location metrics M1 and M2 values didn't improve as the budget was increased - neither for the regular SNPE method nor for the surrogate. We do not have an explanation for this result. For the dispersion metrics M3 and M4, the means did improve as a function of the budget, though their variances did not. 

\section{Discussion}

Our work proposes and evaluates computational strategies that can reduce the need for simulator calls when using simulation based inference (SBI). One option is to estimate a surrogate that can be used in lieu of the simulator for many of the runs required in SBI. Another is to improve the representativeness of samples via the method of support points. A third option is to employ both of these methods. Each of these is considered as a modification to the Sequential Neural Posterior Estimator (SNPE)  \cite{snpe-a,snpe-b,snpe-c}.\\

On our test bed of challenging cases from the literature, the results indicate that these methods generally improve the performance of SNPE for the same budget of simulator calls. In particular, the surrogate and mixed methods were more successful in approximating the posterior distributions. Improvements were observed for all the metrics that we considered. The frequency and the extent of the improvements varied across task; and often for the same task they were not consistent across simulation budget size. The latter could just be a reflection of the limited testing done for each task by budget combination. We preferred to cover a wide range of tasks and budgets rather than to do extensive testing of a small collection. For the surrogate method, the max improvement in the C2ST was 13.45\% (2-Moons task with a budget of 300, median improvement over 20 runs) and the average across all tasks and budgets was 1.5\%. Lower improvements were observed in the other methods. \\ 

Problems were sometimes encountered with all the methods when training the NDEs. These are complex neural nets and the weights did not always converge to a good fit when applied to the training data. As the surrogate method requires fitting two NDEs, it may be especially susceptible to unsuccessful training. Future research that improves the training of NNs will benefit all of the methods studied here, and perhaps especially our surrogate method.\\

On all the synthetic tasks, the posterior approximations improved with increases in the size of the simulator budget. However, for the HH simulation, the results did not show any consistent improvement. Further research would be helpful to understand why this occurs. \\

Although this work focused on reducing simulator calls, our proposed method of using two NDE's side by side (one for the likelihood and one for the posterior) might also be an alternative to the SNLE algorithm. Traditional SNLE uses an NDE to approximate the likelihood and then switches to MCMC to obtain samples from the posterior. We can instead treat the likelihood NDE as a surrogate, use it to sample additional data points, and then train a posterior NDE on the whole data. This can be used to replace the MCMC/VI step completely, or interwoven between different rounds of SNLE (the posterior NDE can replace the MCMC/VI). And this can be done by spending all the simulator budget on the 1st round or distributing it across rounds where the true simulator is used (the likelihood NDE can replace the simulator). This idea was only partially explored in this work. 

\section{Acknowledgements}
This research has received funding from the European Union’s Horizon 2020 Framework
Programme for Research and Innovation under the Specific Grant Agreement 
No. 945539 (Human Brain Project SGA3).

\clearpage

\appendix

\section{Appendix A: Tasks}

Here we'll expand on the test problems (tasks) used to test and compare the algorithms. The test problems employed in the article have all been published elsewhere. We provide here a detailed description of each problem.\\

\begin{itemize}
    \item \textbf{Gaussian Mixture Model (GMM) with same mean and different variance (1D)} \cite{snpe-a} - The true posterior is the following GMM: $p(\theta|x) = 0.5\phi(\theta;x,0.1^2)+0.5\phi(\theta;x,1^2)$ where $\phi$ is the Normal PDF. This distribution has a long non-Gaussian tail. One way of getting to this posterior is to have the same GMM likelihood $p(x|\theta) = 0.5\phi(x;\theta,0.1^2)+0.5\phi(x;\theta,1^2)$ and a wide enough uniform prior $p(\theta)\propto 1$ such that in the posterior $x$ and $\theta$ will simply switch roles. In this task the prior was $\theta\sim U(-10,10)$, and $x_{obs}=0$. Note that technically the posterior is a truncated GMM which has to be normalized. But given that $x_{obs}=0$, the truncation is insignificant.
    \\
    \item \textbf{Bayesian Linear Regression (6D)} \cite{snpe-a} - Prior is a standard 6D multivariate normal $p(\theta) = \mathcal N(0,I)$. The covariates of the Linear Regression, here denoted as $X$, were sampled also from a standard multivariate normal, but this was performed 10 times, so $X$ is a ($10\times6$) matrix. Finally $y|\theta\sim \mathcal N(y; X\theta, \sigma^2\cdot I)$. So $y$ is our (10D) data. The posterior is also a (multivariate) Gaussian. Let us find its precision and mean:
    $$p(y|\theta) = \mathcal N (X\theta, \sigma^2\cdot I), p(\theta) = \mathcal N(0, I)$$
    $$ p(\theta|y) \propto  p(y|\theta)\cdot p(\theta) \propto e^{-0.5(y-X\theta)^T(\sigma^2\cdot I)^{-1}(y-X\theta)}\cdot e^{-0.5\theta^T\theta} $$
    $$\propto e^{-0.5[\frac{1}{\sigma^2}(\theta^TX^TX\theta-2\theta^TX^Ty)+\theta^T\theta]}= e^{-0.5[\theta^T(\frac{1}{\sigma^2}X^TX+I)\theta-2\frac{1}{\sigma^2}\theta^TX^Ty)]}
    $$
    This implies that the posterior is also of Gaussian form, with a precision $\Sigma^{-1}=\frac{1}{\sigma^2}X^TX+I$ and a mean $\mu=\frac{1}{\sigma^2}\Sigma X^Ty$. 
    As in \cite{snpe-a} we took $\sigma=0.1$. \\
    
    \item \textbf{Two Moons (2D)} \cite{snpe-c} - It is a two-dimensional posterior that exhibits both global (bimodality) and local (crescent shape) structure. The prior is 2D box-uniform $\theta\sim U(-1,1)$. The simulator is defined as follows:
    $$ x|\theta = \begin{bmatrix}
           r \cos \alpha + 0.25 \\
           r \sin \alpha \\
         \end{bmatrix} + \begin{bmatrix}
           - |\theta_1+\theta_2|/\sqrt{2} \\
           (- \theta_1+\theta_2)/\sqrt{2} \\
         \end{bmatrix},  \alpha \sim U(-0.5\pi, 0.5\pi),  r\sim\mathcal N(0.1, 0.01^2).
    $$
    The terms $\alpha$ and $r$ represent the internal stochasticity of the simulator. Let us denote the 1st vector in the above sum as $v$. Note that $v$ has a distribution concentrated around an arc of radius 0.1 (``crescent shape''). The absolute value in the 2nd vector causes the bimodality in the posterior. Finding the analytical likelihood is very challenging. But we can generate samples from the posterior rather simply. Since the simulator relates to the parameters, we could invert the equation to get the parameters as a function of the observed $x$: i.e., suppose we have a single observation $x_{obs}=(0,0)$, then we could generate/sample different $v$'s and then invert the equation:
    $ x = v + (-\frac{|\theta_1+\theta_2|}{\sqrt{2}},\; \frac{-\theta_1+\theta_2}{\sqrt{2}})^T
    $ to get samples from the posterior.
    Because of the absolute value, we get 2 different solutions and we can simply choose one of them randomly.  
    $$v_1 - \frac{|\theta_1 + \theta_2|}{\sqrt2}=x_1, \;\; v_2 + \frac{-\theta_1 + \theta_2}{\sqrt2}=x_2.$$
    Moving sides and defining the right-hand-side as $q$'s:
    $$\frac{|\theta_1 + \theta_2|}{\sqrt2}=v_1-x_1:=q_1, \; \frac{\theta_1 + \theta_2}{\sqrt2}= \pm q_1, \;\; \frac{-\theta_1 + \theta_2}{\sqrt2}=x_2-v_2:=q_2. $$
    
    Subtracting/adding to cancel the variables we get:
    $$\sqrt2 \theta_2=\pm q_1 + q_2, \;\; \sqrt2\theta_1=\pm q_1 - q_2 \Rightarrow \theta_2 = (\pm q_1 + q_2)/\sqrt2, \theta_1 = (\pm q_1 - q_2)/\sqrt2
    $$
\\
    \item \textbf{Simple Likelihood Complex Posterior (SLCP) (5D)}  \cite{snle-b} - a problem specifically designed to have a simpler likelihood than a posterior (i.e., the $x|\theta$ structure is simpler than the $\theta|x$ structure). The prior is uniform on a 5D box $\theta \sim U(-3,3)^5$. The likelihood in the original problem had four 2D points sampled from a Normal distribution, but we chose to take eight points in hope to simplify inference: $x|\theta=(x_1,...,x_8), x_i\sim\mathcal N(m_\theta,S_\theta)$, where:
    $$m_\theta = \begin{bmatrix}\theta_1 \\ \theta_2 \end{bmatrix}, S_\theta = \begin{bmatrix}s_1^2 & \rho s_1 s_2 \\ \rho s_1 s_2 & s_2^2 \end{bmatrix}, s_1 = \theta_3^2, s_2=\theta_4^2, \rho = \tanh \theta_5$$ 
    In practice, $x$ is flattened into a 16D vector when fed into the SBI algorithm. We took the same ground truth parameters $\theta_{true} = (0.7,-2.9,-1,-.9,0.6)$ as \cite{snle-b}. Despite the model’s simplicity, the posterior is complex and non-trivial: it has four symmetric modes (due to squaring), and vertical truncation (due to the uniform prior [which \textbf{IS} significant here]). 
  \\  
    \item \textbf{Bernoulli GLM (10D)} \cite{snpe-b} -  Generalized linear models (GLM) are commonly used to model neural responses to sensory stimuli. Neural activity is subdivided into ``bins'' and, within each bin $k$, spikes are generated according to a Bernoulli observation model: $y_k\sim Ber(p)$. The canonical logit link function was used, that is, $\log(\frac{p}{1-p})=x_k^T\theta$. $x_k$ consists of an intercept term (constant 1) and 9D sliding window (the ``bin'') on a series of stimuli (Gaussian white noise) of size $T=100$. The initial window is padded with $0$'s. The design matrix $X$ is hence of shape $(100\times10)$, and $\theta$ is 10D. The prior on $\theta$ has a covariance matrix which encourages smoothness by penalizing the 2nd-order differences in the vector of parameters, that is $\theta \sim \mathcal N (0, \Sigma)$, where $\Sigma_{1,1}=2$, and $\Sigma_{2:10,2:10}=(F^TF)^{-1}$. $F$ is defined as follows: for $j=1,...,9: F_{j,j-2}=1, F_{j,j-1}=-2, F_{j,j} = 1+\sqrt{\frac{j-1}{9}}$. All other terms (in $F$ and in $\Sigma$) are $0$. While $y$ is our $100\times 1$ data, sufficient summary statistics are used $S:=X^Ty$, a $10\times 1$ vector. Thus, for each parameter vector $\theta_i$ we have a data vector $y_i$ and a summary vector $S_i$. The true posterior sample is found using Pólya-Gamma MCMC \cite{polya}. The true parameters were taken to be as in \cite{snpe-b}: $\theta_{true}=(0.955,-0.452,0.223,1.105,0.271,-0.358,-0.672,-0.206,0.306,-0.436)$.
    \\
    \item \textbf{Sisson's $2^D$ GMM} \cite{sisson} - a toy problem to test the ``Curse of Dimensionality''. We have a GMM with increasing dimensionality: for $D=1$ we have two components in the GMM, for $D=2$ we have four, etc. We took $D=3$. All Gaussians are symmetric around the origin, with a mixing coefficient chosen to be $\omega=0.7$. The problem is set in such a way such that the marginal for each dimension is the same as in the 1D case. A way to write this mathematically is:
    $$ p(x|\theta) \sim \sum_{b_1=0}^1 ... \sum_{b_D=0}^1 [\prod_{i=1}^D \omega^{1-b_i}(1-\omega)^{b_i}]\mathcal N(\mu, \Sigma) $$
    $$\mu = [(1-2b_1)\theta_1,...,(1-2b_D)\theta_D] $$
    In our problem we took $x_{obs}$ to be 1 observation, a vector of $5$'s, and $\Sigma$ to be a matrix with $1$ on the diagonal and $0.7$ correlation for all off-diagonal terms. The prior is uniform on a 3D box with $\theta\sim U(-20,40)^3$. In the posterior $x$ and $\theta$ simply switch roles. Note that technically the posterior is a truncated GMM which has to be normalized (though the truncation is insignificant).\\
\end{itemize}

Note that although increasing the dimensionality usually increases the complexity of the inference, there are also other factors at play. For example, Sisson's 3D task was usually more challenging for the algorithms to infer well, compared to the 6D Bayesian Linear Regression or the 10D Bernoulli GLM - probably due to it having multiple modes, instead of the single mode of the higher dimension problem. \\

In addition we also tested the following:\\

\begin{itemize}
    \item \textbf{Compare the Surrogate method to SNLE} - we used both the surrogate method and the SNLE method on the 5D-SLCP toy problem. SNLE uses an NDE to approximate the likelihood and then continues using MCMC to obtain samples from the posterior. We instead used the likelihood NDE as a surrogate and sampled more data points using it, before training a posterior NDE on the whole data. The SLCP problem was chosen as it is a problem whose likelihood structure is simpler and where SNLE is supposed to preform better, and we wanted to test how the surrogate method preforms compared to it. 
\\
    \item \textbf{Hodgkin-Huxley model (8D)} - this simulator model describes action-potential in neurons. The problem was 1st used in SBI in \cite{snpe-b} but with 10D parameters, though later presented in \cite{eLife} with 8D - that is the version that we also implemented. The simulator creates an electrical trace of ``action potential''. The trace is then summarized into 13D summary statistics: the number of spikes; 5 auto-correlations (starting at 0.1-0.5 seconds from the action potential); the resting potential mean, its standard-deviation, and the mean current during action potential; and the 4 centralized moments (2nd, 3rd, 4th and 5th order) of the action potential trace (We used 13 dimensions instead of the 7 used in \cite{eLife}). There are 8 unknown conductance parameters that we wish to infer, the rest are presumed fixed/known. The prior distribution over the parameters is uniform and centered around the true parameter values: $\theta \sim U(0.5\theta^*, 1.5\theta^*)$. We used the same true parameter as in \cite{eLife}, namely: $(g_{Na}, g_K, g_l, g_M, T_{max}, -V_{T}, \sigma, -E_l) = (50, 5, 0.1, 0.07, 600, 60, 0.1, 70)$.
\end{itemize}

\section{Appendix B: SNLE Results}
We compared the SNLE method to the surrogate method on one selected task: the 5D SLCP. This problem was chosen as it is one which favors the SNLE method (as the likelihood structure is simpler than the posterior structure). Sampling from the posterior in SNLE follows with either MCMC or VI, and the default in the SBI library is MCMC, which we didn't change. Sampling with MCMC can be slow, and so we wanted to compare this method to a surrogate method which might be faster to sample. We compared two types of NDE's: Mixture Density Network (MDN) and Neural Spline Flows (NSF). MDN, although possibly less expressive than NSF, allows for faster sampling from the posterior - as it outputs a normalized distribution, unlike NSF, whose output is non-normalized. We also compared to a surrogate mix of MDN and NSF, where the surrogate is trained with an MDN (for faster sampling) and the posterior is trained with NSF. \\

All methods used 2 rounds of SNLE or of the Surrogate method, with 10 repeated experiments per each budget. The budgets were 250, 500, 1000, 2500, 5000, 7500, 10000. We ran the experiments twice: one run took $5n$ samples from the surrogate, and the other took $10n$ samples. \\

\begin{itemize}
    \item \textbf{MDN} - the surrogate with the $5n$ sampling showed good results only for 2 budgets: 250 and 2500. The rest of the budgets showed bad results: 500, 1000, 5000, 7500, 10000.  The 3 largest budgets had significantly worse results. So, overall significantly bad results. The picture is even less favorable when looking at medians and IQRs: all budgets have bad results. The surrogate version was however usually faster, with (mean) time differences of 14.0879, 7.9748, 1.8930, -6.1686, 122.3127, 131.8625 and 317.7489 seconds for the matching budgets. With the $10n$ sampling, the surrogate method had good results for 250, 500, 1000 and 2500 budgets, but bad results for the 5000, 7500 and 10000. The improvements were more substantial. The good:bad ratio favored the SNLE for MMD and C2ST, but favored Surrogate for ED. Again, the main improvements were found in the 4 smaller budgets; the 3 larger budgets favored the SNLE. The SNLE method was faster than the surrogate across the board, with (mean) time differences (seconds): -5.5, -35.9, -126.8, -553.1, -2018, -2157, -3110. 
\\
    \item \textbf{NSF} - the $5n$ surrogate had good results in only two budgets: 500, 1000. The $10n$ surrogate had good (perfect 9/9) results in only three: 250, 500 and 1000. Surprisingly, the $10n$ had overall greater improvements than performance losses, but the average of good vs. bad rounds favored the SNLE. Looking at medians and IQRs turns the 10000 budget into a good result. The SNLE method was significantly faster.  
\\
    \item \textbf{NSF mixed with MDN} - here we used NSF for the posterior and MDN for the surrogate. Compared to SNLE with NSF the results here were quite bad across the board: no good results for the $5n$ sample and only one good result for the $10n$ sample: the 250 budget. Although using MDN made the surrogate sampling faster, it was still usually slower than SNLE. Compared with SNLE with MDN - the results were quite good, especially for lower budgets: for $5n$ 250, 500, 1000, 2500, 5000 budgets gave good results; for $10n$ 250, 500, 1000, 2500 gave good results. Improvements dominated performance losses in both, and good:bad ratio was better for MMD and ED; C2ST favored the SNLE with $5n$ samples, with a tie in the $10n$ comparison. SNLE with MDN was significantly faster.\\
\end{itemize} 

Note that the main reason the SNLE method is usually faster than the surrogate is the increased sample size ($5n$ or $10n$) with the latter. Otherwise, the surrogate method is expected to be faster on most if not all experiments.\\

Overall, the NSF methods showed much better results than the MDN methods.

\clearpage

\bibliographystyle{plain}
\bibliography{ref}

\end{document}